\documentclass[iop]{emulateapj}

\usepackage{natbib,epsfig,siunitx,url,graphicx,amsmath,footnote,float,xcolor,cases}
\usepackage[colorlinks=true,linkcolor=blue,citecolor=blue]{hyperref}

\begin{document}

\submitted{MNRAS Letters; accepted}

\title{A record of the final phase of giant planet migration fossilized in the asteroid belt's orbital structure}

\author{Matthew S. Clement\altaffilmark{1,2,*}, Alessandro Morbidelli\altaffilmark{2}, Sean N. Raymond\altaffilmark{2}, \& Nathan A. Kaib\altaffilmark{4}}

\altaffiltext{1}{HL Dodge Department of Physics Astronomy, University of Oklahoma, Norman, OK 73019, USA}
\altaffiltext{2}{Department of Terrestrial Magnetism, Carnegie Institution for Science, 5241 Broad Branch Road, NW, Washington, DC 20015, USA}
\altaffiltext{3}{Observatoire de la C{\^o}te d'Azur, BP 4229, 06304 Nice Cedex 4, France}
\altaffiltext{4}{Laboratoire d'Astrophysique de Bordeaux, Univ. Bordeaux, CNRS, B18N, all{\'e} Geoffroy Saint-Hilaire, 33615 Pessac, France}

\altaffiltext{*}{corresponding author email: matt.clement@ou.edu}

\begin{abstract}
The asteroid belt is characterized by an extreme low total mass of material on dynamically excited orbits.  The Nice Model explains many peculiar qualities of the solar system, including the belt's excited state, by invoking an orbital instability between the outer planets.  However, previous studies of the Nice Model's effect on the belt's  structure struggle to reproduce the innermost asteroids' orbital inclination distribution.  Here, we show how the final phase of giant planet migration sculpts the asteroid belt, in particular its inclination distribution.  As interactions with leftover planetesimals cause Saturn to move away from Jupiter, its rate of orbital precession slows as the two planets' mutual interactions weaken.  When the planets approach their modern separation, where Jupiter completes just short of five orbits for every two of Saturn's, Jupiter's eccentric forcing on Saturn strengthens.  We use numerical simulations to show that the absence of asteroids with orbits that precess between 24-28 arcsec yr$^{-1}$ is related to the inclination problem.  As Saturn's precession speeds back up, high inclination asteroids are excited on to planet crossing orbits and removed from the inner main belt.  Through this process, the asteroid belt's orbital structure is reshaped, leading to markedly improved simulation outcomes.
\end{abstract}

\section{Introduction}

As the giant planets grow within the primordial gas disk, the combination of the Sun's radial force and gravitational torques from the disk and other planets rapidly shepherd them into a mutual resonant configuration \citep{masset01,morbidelli07}.  This scenario is consistent with the number of resonant giant exoplanets discovered  \citep[eg: GJ 876 and HR 8799, among others,][]{mills16}.  After the nebular gas dissipates, the giant planets' orbits continue to evolve via interactions with leftover planetesimals in the primordial Kuiper Belt.  Scattering events between these objects and the outermost gas giants preferentially displace material inward, while Jupiter tends to scatter objects out of the system entirely \citep{fer84}.  These small exchanges of angular momentum cause the giant planets' orbits to diverge, eventually destroying the resonant chain.  The Nice Model describes how the global instability induced by this escape from resonance sculpts the primordial solar system into its modern form \citep{Tsi05,mor05,gomes05,nesvorny12}.

The precise timing of the instability has been the subject of a number of recent studies.  A delayed instability would imply a correlation with the late heavy bombardment (a perceived spike in lunar cratering $\sim$400 Myr after gas disk dispersal), the existence of which is now in doubt \citep{zellner17,morb18,quarles19}.  Furthermore, simulations of the Nice Model's effects on the fully formed inner planets routinely over-excite the fragile terrestrial worlds on to orbits where they collide with one another or are lost from the system \citep{bras13,kaibcham16}.  The ''Jumping Jupiter'' \citep{bras09,roig16} model attempts to resolve this issue by requiring that Jupiter and Saturn's semi-major axes diverge in a step-wise manner towards their modern locations as the result of a close encounter with one of the ice giants.  However, studies of the scenario argue for weaker instabilities \citep{deienno18} and large period ratio jumps \citep{toliou16} that are low probability outcomes of statistical studies of the instability \citep{nesvorny12,deienno17}.  Furthermore, authors must post-process simulation results by modifying the asteroid belt's initial inclination distribution in order to provide good matches to the modern orbital structure \citep{roig15,deienno16,deienno18}.  Recent work argued that an early (just a few Myr after gas dissipation) instability \citep{nesvorny18} might fix the terrestrial destabilization problem \citep{clement18,clement18_frag} without requiring such a specific jump.  Such an evolutionary model has the advantage of providing a natural explanation for the disparity between the inferred geological accretion time-scales of Earth \citep[$\gtrsim$50 Myr;][]{kleine09} and Mars \citep[$\lesssim$ 5 Myr;][]{Dauphas11}. However, Xenon measurements from Comet 67P \citep{marty17} are at odds with an instability occurring before the end of Earth's magma ocean phase.  Other successful models \citep[eg:][]{levison15,izidoro15} for terrestrial evolution resolve problems related to Mars' size and formation time by invoking non-uniform disk conditions.  In particular, the ``Grand Tack'' model \citep{walsh11,walsh16} resolves the small Mars problem by arguing that Jupiter migrated into the terrestrial region during the nebular gas phase; thereby truncating the distribution of embryos and planetesmials near Mars' modern orbit. However, all schemes require a  giant planet instability at some time to explain the outer solar system \citep[see recent review in ][]{ray18_rev}.

While the Nice Model is successful at explaining most aspects of the solar system's dynamical state, studies of its consequences in the asteroid belt are all plagued by a common pitfall \citep{obrien07,deienno16,deienno18,clement18_ab}.  Specifically, numerical simulations overpopulate the high inclination parameter space in the inner main belt (we define the inner belt as the region of asteroids with semi-major axes less than 2.5 au).  In this letter, we examine the dynamical processes responsible for this shortcoming.  Additionally, we propose a mechanism through which these high-inclination asteroids are naturally removed that is compatible with any of the various terrestrial evolutionary models.

\section{The Asteroid Belt inclination problem}

The spatial orientation of orbits in the solar system precess circularly, or rotate, on time-scales much longer than their actual orbital periods.  The perturbative effects of these variations within the Keplerian problem (particularly those of the Jupiter-Saturn system) have long been known to drive dynamics in the asteroid belt \citep{poincare1892,morby91}.  In the secular theory of solar system evolution \citep[eg:][]{milani90,morby91,dermott99}, the long-term behavior of the eight planets' orbital eccentricities ($e_{i}$) and longitudes of perihelia ($\varpi_{i}$) are described by the solutions to the secular equations of motion:

\begin{equation}
    \begin{array}{l}
    e_{i}\cos{\varpi_{i}} = \sum_{j}^{8}M_{ij}\cos{(g_{j}t + \beta_{j})} \\
    e_{i}\sin{\varpi_{i}} = \sum_{j}^{8}M_{ij}\sin{(g_{j}t + \beta_{j})}
	\label{eqn:sec}
	\end{array}
\end{equation}

 The same analysis can be applied to the precession of the planets' inclination nodes; specifically the behavior of the orbital inclination ($\sin{i/2}$) and longitude of ascending node ($\Omega$).  Secular resonances occur when an object precesses at a rate equal to one of the solar system's dominant eigenfrequencies.  These eigenfrequencies are denoted $g_{1}$-$g_{8}$ for the planets' eccentricity vector precessions, and $s_{1}$-$s_{8}$ for the inclination node precessions.  The $\nu_{6}$ resonance is comprised of the orbital semi-major axes and inclinations with precession rates that match the $g_{6}$ rate of 28.22 arcsec yr$^{-1}$, and cuts across the modern asteroid belt (see Fig.~\ref{fig:inc}). There is a clear deficiency of inner belt asteroids with inclinations above the $\nu_{6}$ resonance and precessions slower than $g_{6}$ relative to Nice Model predictions \citep{morby10}.  In the modern asteroid belt, the ratio of large ($D>$ 50 km) asteroids with inclinations above the $\nu_{6}$ resonance, to those below is $\sim$0.08 (referred to in the subsequent text as the $\nu_{6}$ ratio).
 
 In the current version of the Nice Model, the orbital eccentricities and inclinations of asteroids are excited by secular resonances rapidly moving across the belt region \citep{obrien07,deienno18}.  Because simulations of the instability begin with the giant planets in a more compact configuration, their orbits precess at different rates than they currently do. The locations of their respective secular resonances are displaced as the planets' orbits change during, and after the instability.  Of most importance for the inner asteroid belt, the $\nu_{6}$ resonance must traverse from $\sim$4.5 au to its modern location at $\sim$2.05 au, and the $\nu_{16}$ (inclination nodes precessing equal to $s_{6}$)  resonance must sweep from $\sim$2.8 au to $\sim$1.9 au \citep{walshmorb11}.  The $\nu_{16}$ resonance moves inward, and encounters the inner belt first.  This process excites inclinations, but leaves the eccentricities of asteroids unaffected.  Because of the characteristic shape of $\nu_{6}$ (Fig ~\ref{fig:inc}), its movement only excites the eccentricities of low-inclination  asteroids in the inner belt (often on to planet crossing orbits).  Through these processes, many asteroids are stranded on relatively stable, high-inclination, low-eccentricity orbits in the inner main belt \citep{morby10,clement18_ab}.  
 
 The fraction of inner main belt asteroids isolated above $\nu_{6}$ is tied to the smoothness of the giant planets' migration.  Studies of smooth migration utilizing artificial forces substantially deplete the $a/i$ parameter space below $\nu_{6}$, and simultaneously overpopulate the high-$i$ region of phase space \citep{morby10,walshmorb11}.  However, lower $\nu_{6}$ ratios can be achieved with a ``Jumping Jupiter'' style instability \citep{bras09,roig15,deienno16,deienno18}, or when the full chaos of the event is considered \citep{clement18,clement18_ab}.  While many inner belt asteroids' inclinations are still over-excited in these scenarios, a substantial number survive giant planet migration with inclinations below $\nu_{6}$ because the important secular resonances do not linger at any particular location \citep{clement18}.

\section{Depletion Mechanism}
Dynamical instabilities are inherently stochastic, and each follows a unique path.  When studying the Nice Model, authors typically select systems in which the giant planets' final orbits are closest to those of the modern solar system \citep{clement18,deienno18}.  However, this does not guarantee that the simulated planets followed the same evolutionary path as the real ones.  As Saturn's orbit moves away from the Sun following the instability, its precession rate continues to decrease, thereby lowering $g_{6}$ to its modern value.  Thus the crux of the Nice Model's $\nu_{6}$ problem has been in finding a mechanism to deplete asteroids that precess slower than $g_{6}$ after the instability strands them above $\nu_{6}$.  However, these previous studies have neglected the precise effects induced by Jupiter and Saturn's specific modern configuration.  Presently, the solar system's two most massive planets lie just inwards of a mutual 5:2 mean motion resonance (MMR), with Jupiter completing $\sim$4.97 orbits for every two of Saturn's.  

\begin{figure}
\centering
\includegraphics[width=.5\textwidth]{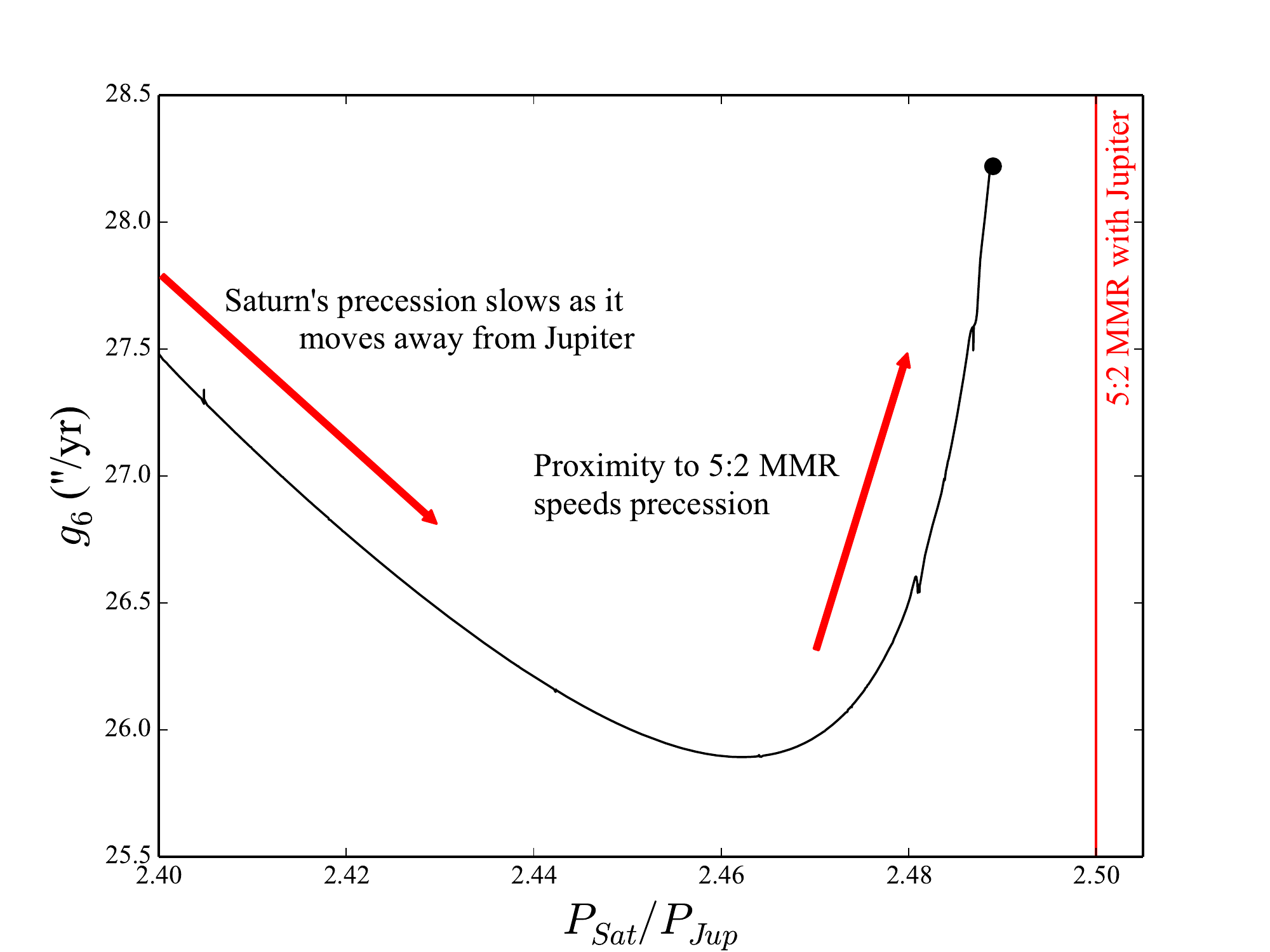}
\caption{Evolution of the solar system's $g_{6}$ eigenfrequency during Saturn's final phase of migration.  The modern value of $g_{6}$ is denoted by a bold point. The figure's minimum is at $P_{Sat}/P_{Jup}=$ 2.46 and $g_{6}=$ 25.89.}
\label{fig:g6}
\end{figure}

Secular precessions are known to speed-up near MMRs \citep{milani90,morby91}.  Fig.~\ref{fig:g6} demonstrates the behavior of the $g_{6}$ rate as Jupiter and Saturn approach their modern configuration.  Perturbative derivations of the three-body secular Hamiltonian typically expand the problem in Taylor series, and subsequently neglect mass terms of order two or higher \citep[for a full discussion of secular resonances in the asteroid belt see][]{milani90,morby91,morby91b}.  This simplification holds when the bodies' mean longitudes ($\lambda_{i}$ in Delaunay variables) are non-resonant.  When the two objects approach a MMR, the quadratic mass term is no longer negligible, and the precession rate $g_{i}$ increases asymptotically.  Our first set of simulations are designed to measure the solar system's $g_{6}$ eigenmode as Jupiter and Saturn approach the 5:2 MMR.  We perform 3,200 integrations of the modern solar system with the $Mercury6$ hybrid integrator \citep{chambers99}.  In each run, Saturn's semi-major axis is decreased by 0.005 au, and all other orbital elements are left unchanged.  Each system is integrated for 10 Myr, and the secular amplitudes and frequencies are calculated via Fourier analysis of the simulation time outputs \citep{nesvorny96}.  Through this process, we generate the curve presented in Fig.~\ref{fig:g6}.  

\begin{figure}
\centering
\includegraphics[width=.5\textwidth]{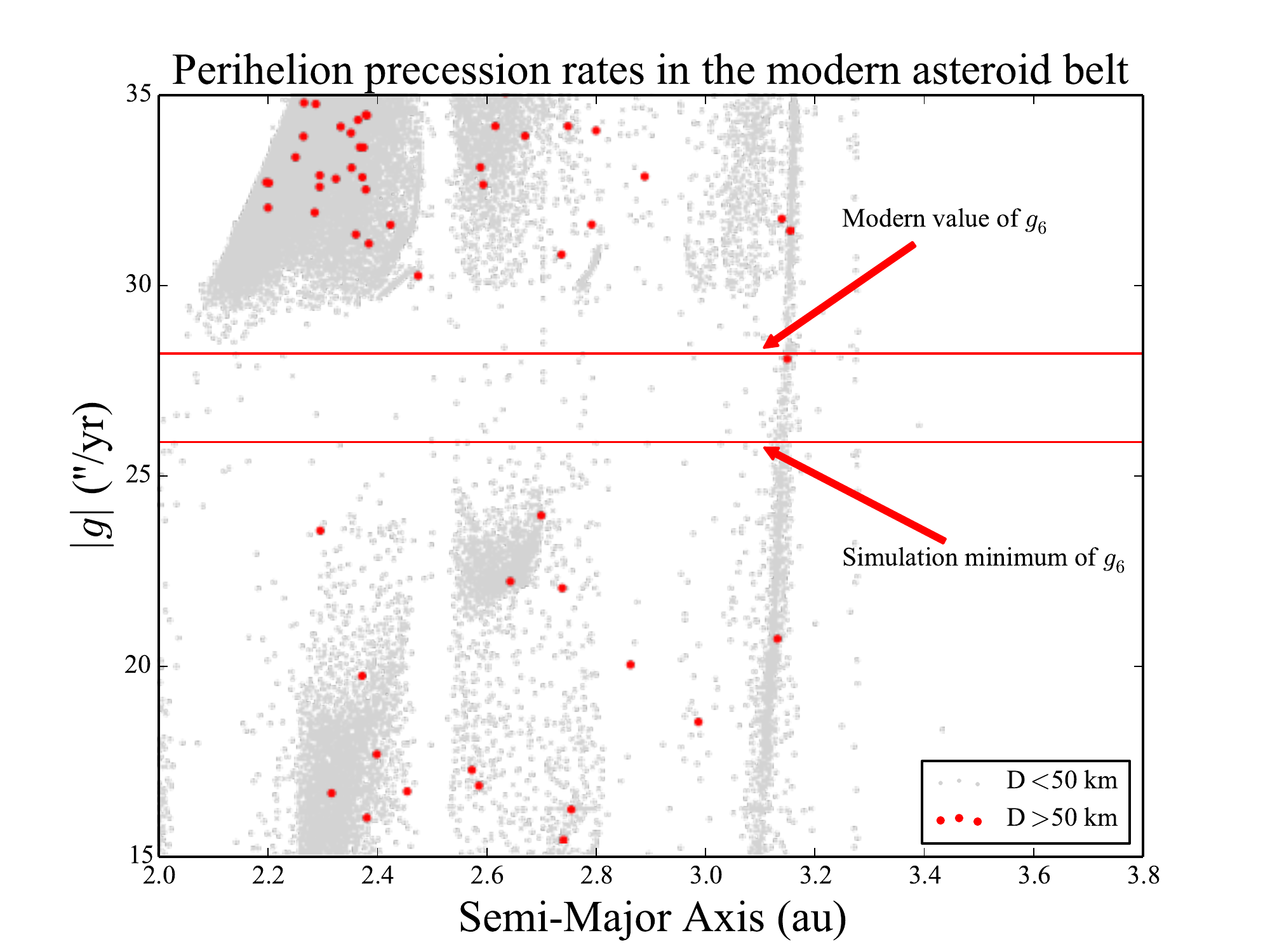}
\caption{Distribution of orbital precession rates as a function of semi-major axis for all known asteroids with constrained orbits \citep{ast_dys}.  The horizontal lines represent the current value of Saturn's $g_{6}$ eigenfrequency, and the minimum value obtained from Fig.~\ref{fig:g6}.  Red points correspond to large asteroids for comparison with the bottom panel of Fig.~\ref{fig:inc} (note that the total number of points is less here as this figure is zoomed in on the range of $15<g<35$).  The asteroids in between the red lines near $\sim$3.1 au are members of the high-inclination collisional family (31) Euphrosyne \citep{novakovic11}.  After the break-up of Euphrosyne below the bottom red line, the family members filled the gap that was presumably emptied by primordial migration as the result of semi-major axis spreading due to the Yarkovsky Effect \citep[eg:][]{bottke01}.}
\label{fig:g_ab}
\end{figure}

As the value of $g_{6}$ lowers and rises, the $\nu_{6}$ resonance shape sweeps from right to left and back in $a/i$ space.  We argue that this reversal in sweeping of the $\nu_{6}$ resonance explains the depletion of asteroids with precession rates less than the current value of $g_{6}$ (Figs.~\ref{fig:g_ab} and ~\ref{fig:g_sim}) and inclinations above $\nu_{6}$ in the inner belt region (Fig.~\ref{fig:inc}, top panel).

\section{Numerical Simulations}

Next, we use numerical simulations to demonstrate how the reversal of $g_{6}$ by just $\sim$2.5 arcsec yr$^{-1}$ can affect a uniform population of main belt asteroids about the $\nu_{6}$ secular resonance (migrating the Jupiter-Saturn period ratio from $\sim$2.45-2.49).  We use the $GENGA$ \citep{genga} integration package for this phase of our study.  We first test different migration time-scales by performing three separate simulations of the solar system and 10,000 massless test particles.  Asteroid orbital elements are selected randomly from uniform distributions of non-planet-crossing orbits (2.0 $<a<$ 4.0 au, 0.0 $<e<$ 0.5, 0.0 $<i<$ 40.0$^{\circ}$ and 0-360$^{\circ}$ for angular orbital elements).  Simply put, the migration is achieved by minor alterations to Saturn's semi-major axis (by cubic interpolation utilizing $GENGA's$ built in \textit{Set Elements} function) such that the Jupiter-Saturn period ratio evolution follows an exponential function of time until reaching the modern value.  Each system is integrated for an additional 100 Myr to remove quasi-stable asteroids.  Our selected migration speeds ($\tau_{mig}$) are loosely based on studies of Saturn's smooth migration's effect on the asteroid belt's structure \citep[as we seek to study uniform migration after the Nice Model instability;][]{minton11}.  Specifically, our fastest migration ($\tau_{mig}=$3 Myr) is selected to equal the slowest sweeping of $\nu_{6}$ (through the bulk of it's migration from 2.8-2.1 au) that permits the asteroid belt's survival.  Because we are only interested in the final phase of migration and clearing in the young solar system, our selected migration speeds are quite slow.  

\begin{figure}
	\centering
	\includegraphics[width=.5\textwidth]{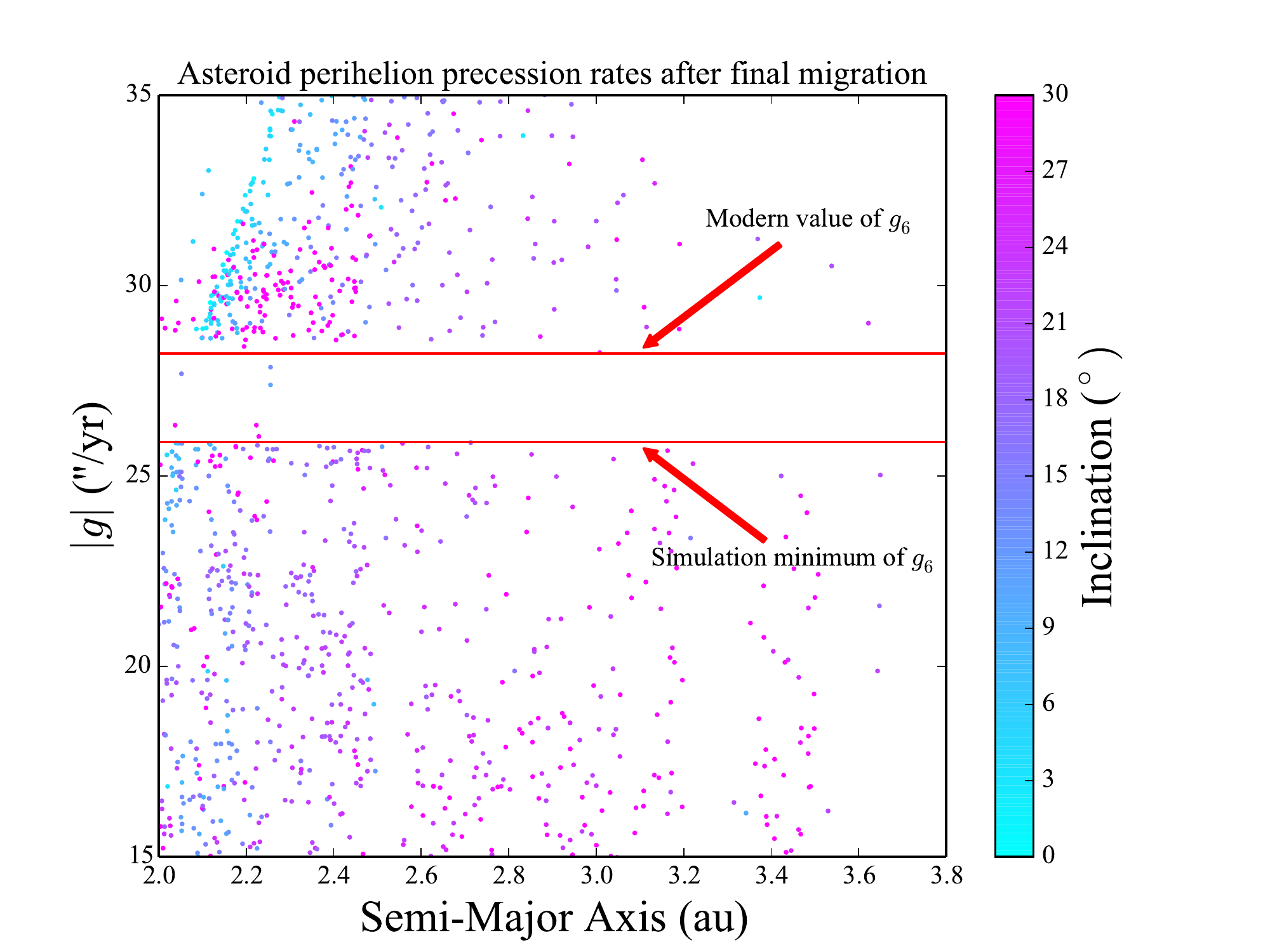}
	\caption{Final asteroid precession rates from our $\tau_{mig}=$ 30 Myr simulation of a uniform distribution of asteroids (table \ref{table:uni}, run 3).  Precession rates are computed by frequency modulated Fourier Transform \citep[see:][]{nesvorny96,ast_dys}. The colour of each point corresponds to the object's inclination.  Note that, as with Fig. ~\ref{fig:g_ab}, the total number of points here is not 10,000 since many asteroids precess faster or slower than the range of values plotted (particularly in the outer main belt).}
	\label{fig:g_sim}
\end{figure}

The results of these simulations are summarized in Table~\ref{table:uni} and Fig. ~\ref{fig:g_sim}.  Through the full migration process, the $\nu_{6}$ ratio consistently drops from 1.98 to less than unity.  Asteroids with high inclinations that would have been unaffected if $g_{6}$ had never dipped below $\sim$28 arcsec yr$^{-1}$ are quickly swept up and elevated in eccentricity via resonant perturbations (Fig. ~\ref{fig:g_sim}).  Once excited, they are eventually removed from the belt as the result of encounters with the terrestrial planets (largely Earth and Mars).  Through this process, the inner belt's overall $\nu_{6}$ ratio is substantially reduced.  Furthermore, these results are independent of the migration time-scale selected (3.0, 10.0 and 30.0 Myr).  Since Saturn's final migration has a strong effect on the inner main belt's population above the $\nu_{6}$ resonance regardless of migration speed, we limit $\tau_{mig}$ to 5 Myr for the remainder of our study \citep{morby10,toliou16}.

\begin{table}
\caption{Initial conditions and results for simulations of a uniform distribution of asteroids: the columns are as follows: (1) the simulation number, (2) the total simulation time, (3) Saturn's average migration speed for the first 0.04 au, and (4-5) the initial and final ratios of inner main belt ($a<$ 2.5 au, $i<$ 40$^{\circ}$) asteroids above to those below the $\nu_{6}$ resonance.}
\label{table:uni}
\centering
\begin{tabular}{c c c c c}
\hline
Run & $\tau_{mig}$ & $\dot{r}_{Sat}$ & $\nu_{6}$ ratio$_{i}$ & $\nu_{6}$ ratio$_{f}$ \\
\hline
1 & 3 Myr & 0.1 au Myr$^{-1}$ & 1.98 & 0.81 \\
2 & 10 Myr & 0.03 au Myr$^{-1}$ & 1.98 & 0.85 \\
3 & 30 Myr & 0.01 au Myr$^{-1}$ & 1.98 & 0.75 \\
\hline
\end{tabular}
\end{table}
 
As evidenced by the reduction of the $\nu_{6}$ ratio by a factor of three or so, this mechanism is successful at removing asteroids near $\nu_{6}$, but cannot be solely responsible for generating the modern ratio.  Therefore, the ratio must also be limited in Jupiter's jump phase \citep{roig15,deienno16,toliou16,clement18,deienno18,clement18_ab}.  In our next set of simulations, we investigate asteroid belts formed via terrestrial accretion models \citep{clement18} that experienced a range of jumps.  We begin by selecting all surviving asteroids from successful simulations of the Early Instability scenario in \citet{clement18_ab} that finished with $P_{Sat}/P_{Jup}<$2.8 (using the nomenclature of that work these are runs 1, 3, 6, 1b and 2b).  Each system in \citet{clement18_ab} was evolved for 200 Myr, through the Nice Model instability and giant impact phase of terrestrial planet formation.  Because a giant planet instability of arbitrary timing is invoked to explain the outer solar system in all terrestrial planet formation models, our initial conditions can be considered roughly independent of evolutionary scheme \citep{ray18_rev}.  To improve statistics, we generate 10 separate, 1,000-particle belts by randomly choosing asteroids from these completed simulations, and slightly altering their semi-major axes, eccentricities and inclinations.  These small positive and negative deviations are made via random sampling of Rayleigh distributions ($\sigma_{a}=$ 0.025 au, $\sigma_{e}=$ 0.025 and $\sigma_{i}=$ 1.0$^{\circ}$).  All 8 planets, and the three largest modern asteroids are included for these simulations.

\begin{table}
\caption{Results for simulations of asteroid belts generated via 200 Myr planet formation simulations \citep{clement18_ab}: the columns are as follows: (1) the simulation number, and (2-3) the initial and final ratios of inner main belt ($a<$ 2.5 au, $i<$ 40$^{\circ}$) asteroids above to those below the $\nu_{6}$ resonance.}
\label{table:results}
\centering
\begin{tabular}{c c c c c}
\hline
Run & $\nu_{6}$ ratio$_{i}$ & $\nu_{6}$ ratio$_{f}$ \\
\hline
1 & 1.45 & 0.67 \\
2 & 1.07 & 0.0 \\
3 & 1.69 & 1.14 \\
4 & 1.37 & 0.55 \\
5 & 1.81 & 0.33 \\
6 & 2.0 & 0.25 \\
7 & 1.12 & 0.42 \\
8 & 1.14 & 1.0 \\
9 & 1.15 & 0.60 \\
10 & 1.63 & 0.63 \\  	  	  	
\hline
\end{tabular}
\end{table}

We provide the initial and final $\nu_{6}$ ratios for these simulations in Table~\ref{table:results}.  The median initial $\nu_{6}$ ratio for our simulations is $\sim$1.4 (as compared to the modern solar system value of $\sim$0.08).  After 100 Myr of evolution, the overall result of Fig. ~\ref{fig:g_sim} and Table ~\ref{table:uni} holds.  In the two outlier simulation (runs 3 and 8), the sweeping of $\nu_{6}$ destabilized many low-$i$, as well as high-$i$ asteroids.  Therefore, while the inclination parameter space above $\nu_{6}$ was well depleted, the final ratio was still poor.
 
\begin{figure}
\centering
\includegraphics[width=.5\textwidth]{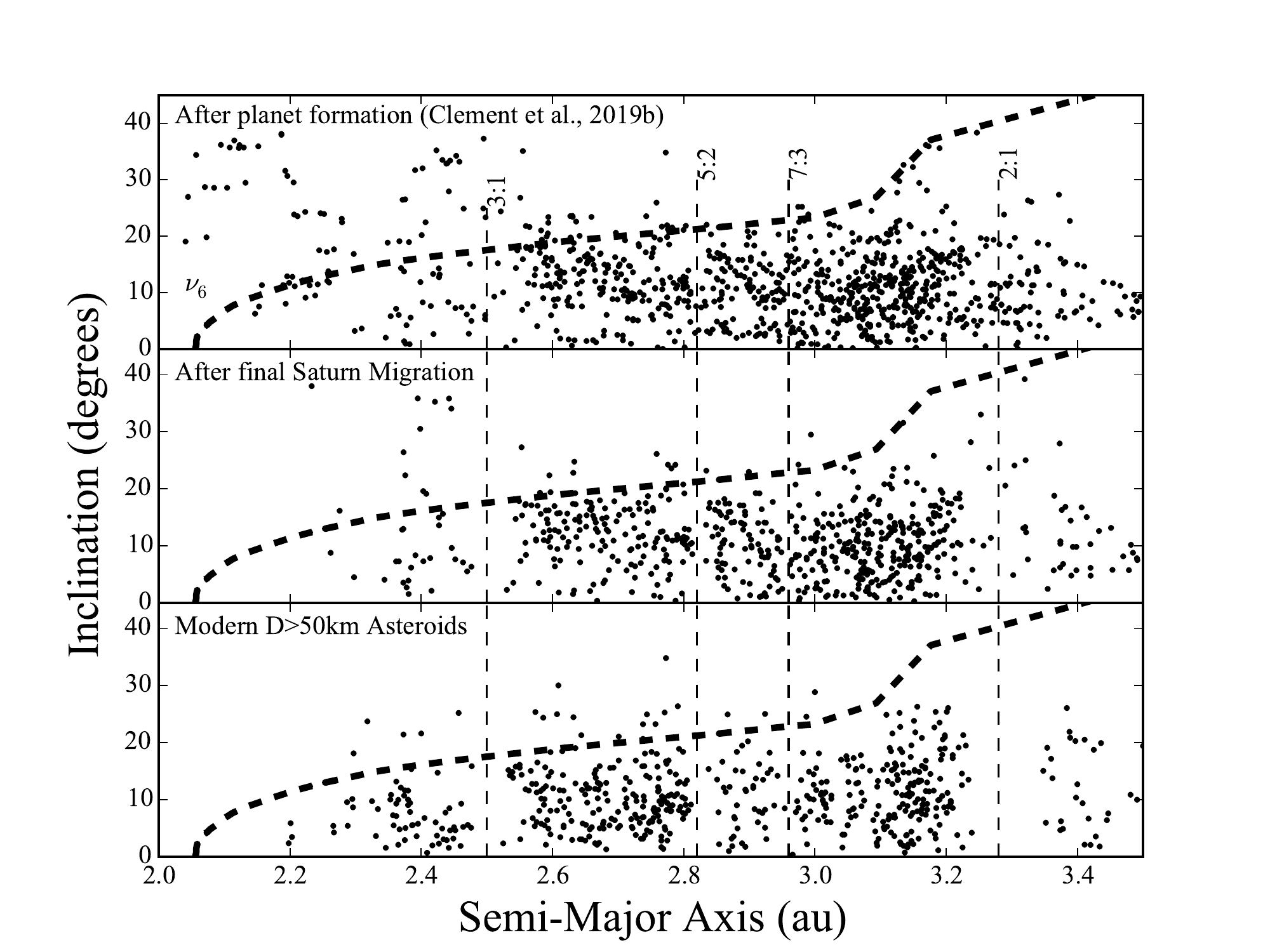}
\caption{Inclination distribution of our simulated asteroid belt compared with the observed structure.  The vertical dashed lines correspond to the semi-major axes of several important mean motion resonances with Jupiter.  The bold dashed lines represent the approximate, eccentricity-averaged orientation of the $\nu_{6}$ secular resonance in $a/i$ space.  The top panel depicts the initial conditions for a successful simulation.  The middle panel illustrates the same belt's inclination structure following 5 Myr of Saturn's final migration inducing a reversal in the sweeping of $\nu_{6}$, and 100 Myr of subsequent evolution in the presence of a steady-state solar system.  The bottom panel shows the present day asteroid belt's $a/i$ distribution (only bright objects with absolute magnitude H $<$ 9.7, approximately corresponding to D $>$ 50 km, are plotted).}
\label{fig:inc}	
\end{figure}

 The inclination structure of a successful simulation (run 5) is plotted in Fig.~\ref{fig:inc}.  While the top panel (the post-planet formation belt) broadly matches the concentrations of modern asteroids in different radial bins, the inner main belt is significantly over-populated above the $\nu_{6}$ resonance.  Contrarily, the middle panel (results of this study) is in better overall agreement with the observed belt (final $\nu_{6}$ ratio of 0.33).  From this figure, it is clear that the sweeping most efficiently removes high-inclination objects in the inner main belt.  Since all objects with $a\lesssim$2.3 au have high inclinations initially (the result of the location and movement of $\nu_{16}$, as discussed above), the region's final $a/i$ structure is altered more dramatically than the 2.3$\lesssim a \lesssim$ 2.5 au region.
 
\section{Conclusions}
The Nice Model \citep{Tsi05,mor05,gomes05,nesvorny12,deienno17} offers the most consistent explanation for the solar system's precise dynamical state.  Numerous authors have investigated the model's effect on the asteroid belt's orbital distribution \citep{morby10,roig15,deienno16,clement18,deienno18,clement18_ab}.  However, previous studies have consistently struggled to match the asteroid belt's inclination population about the $\nu_{6}$ resonance.  We have shown that this is likely resolved when Jupiter and Saturn's precise approach to their 5:2 MMR is considered along with the aforementioned works.  Our current work, coupled with the well developed Nice Model, thus represents a comprehensive picture of the young solar system's formation and early evolution.

\section*{Acknowledgements}
This material is based upon research supported by the Chateaubriand Fellowship of the Office for Science and Technology of the Embassy of France in the United States.  M.S.C. and N.A.K. thank the National Science Foundation for support under award AST-1615975.  S.N.R. acknowledges NASA Astrobiology Institute’s Virtual Planetary Laboratory Lead Team, funded via the NASA Astrobiology Institute under solicitation NNH12ZDA002C and cooperative agreement no. NNA13AA93A.  This research is part of the Blue Waters sustained-petascale computing project, which is supported by the National Science Foundation (awards OCI-0725070 and ACI-1238993) and the state of Illinois. Blue Waters is a joint effort of the University of Illinois at Urbana-Champaign and its National Center for Supercomputing Applications.

\bibliographystyle{apj}
\bibliography{nu6_mnras_arxiv.bib}
\end{document}